\documentclass[prd,aps,showpacs,nofootinbib]{revtex4}
\usepackage{amssymb}
\usepackage{graphicx}
\usepackage{amsmath}

\newcommand{\be}{\begin{equation}}
\newcommand{\ee}{\end{equation}}
\newcommand{\bq}{\begin{eqnarray}}
\newcommand{\eq}{\end{eqnarray}}

\begin{document}
\title{Variation of the speed of light and a minimum speed in the scenario of an inflationary universe with accelerated expansion} 
\author{Cl\'audio Nassif Cruz and Fernando Ant\^onio da Silva}
\affiliation{\small{*CPFT: Centro de Pesquisas em F\'isica Te\'orica, Rua Rio de Janeiro 1186/s.1304, Lourdes, 30.160-041, 
Belo Horizonte-MG, Brazil.\\
  claudionassif@yahoo.com.br, fernando-antonio.silva@caixa.gov.br}} 

\begin{abstract}
In this paper we aim to investigate a deformed relativistic dynamics well-known as Symmetrical Special Relativity (SSR) related to a cosmic background field that plays the role of a variable vacuum energy density associated to the temperature of the expanding universe with a cosmic inflation in its early time and an accelerated expansion for its very far future time. In this scenario, we show that the speed of light and an invariant minimum speed present an explicit dependence on the background temperature of the expanding universe. Although finding the speed of light in the early universe with very high temperature and also in the very old one with very low temperature, being respectively much larger and much smaller than its current value, our approach does not violate the postulate of Special Relativity (SR), which claims the speed of light is invariant in a kinematics point of view. Moreover, it is shown that the high value of the speed of light in the early universe was drastically decreased and increased respectively before the beginning of the inflationary period. So we are led to conclude that the theory of Varying Speed of Light (VSL) should be questioned as a possible solution of the horizon problem for the hot universe.  
 \end{abstract}
\maketitle

\section{Introduction}

The advent of Varying Speed of Light (VSL) theories\cite{1}\cite{2}\cite{3}\cite{4}\cite{5}\cite{6} 
seems to shake the foundations of
Special Relativity (SR) theory, since the speed of light $c$ in vacuum is no longer constant. However, we must take care to investigate 
the veracity of such proposals. To do that, first of all we should consider a new Deformed Special Relativity (DSR) so-called Symmetrical
Special Relativity (SSR)\cite{7}\cite{8}\cite{9}\cite{10}\cite{11}\cite{12}\cite{13}that presents a new causal structure of spacetime
where there emerges an invariant minimum speed $V$ connected to a universal background field given by the vacuum energy, so that we have
$V$ and the speed of light $c$ as being the invariant speeds for lower and higher energies respectively. So in view of SSR theory, this 
paper aims to go beyond by investigating the speed of light $c$ and the universal minimum speed $V$ in the modern cosmological scenario
with a cosmic inflation in the early universe and a very rapid accelerated expansion for a so far future time. Due to the cosmic inflation
at much higher temperatures $T\leq T_P(M_Pc^2/K_B\sim 10^{32} K)$ (Planck temperature) and the rapid accelerated expansion for a far 
future time at much lower temperatures $T\geq T_{mim}(M_PV^2/K_B\sim 10^{-12} K)$\cite{13}, i.e., the minimum temperature related to the 
cosmological horizon, we should take into account an extended SSR with the presence of an isotropic background field with temperature
$T$, where all the particles are moving with respect to a preferred reference frame that plays the role of a universal thermal bath 
with a temperature $T$ decreasing in the time of the expanding universe. Thus we are considering that the energy scale at which a certain
particle is subjected has a nonlocal origin by representing the background thermal energy of the whole universe, i.e., the particle 
should be coupled to the background field with temperature $T$. Such a background thermal effect that leads to a correction on its total
energy ($E=m_0c^2\sqrt{1-V^2/v^2}/\sqrt{1-v^2/c^2}$\cite{11}) is much more pronounced during the inflationary period and a too far future time governed only by vacuum.  

The background thermal effect will allow us to obtain the speed of light with an explicit dependence on the temperature of the universe. So we will be able to preserve the postulate of constancy of the speed of light and extend it just for the implementation of the temperature of the expanding universe. In this sense, we have a function $c(T)$ and so we will find an enormous value for the speed of light in the early universe when its temperature was extremely high close to the Planck temperature $T_P$. In addition, we will also conclude that the very high speed of light was rapidly damped to a value much closer to its current value even before the beginning of the cosmic inflation. This result will lead us
to question VSL theory as an alternative explanation for the horizon problem.  

\section{A brief review of Symmetrical Special Relativity}

The breakdown of Lorentz symmetry for very low energies\cite{10}\cite{11} generated by the presence of a background field is due to an invariant mimimum speed $V$ and also a universal dimensionless constant $\xi$\cite{11}, working like a gravito-electromagnetic constant, namely:

\begin{equation}
\xi=\frac{V}{c}=\sqrt{\frac{Gm_{p}m_{e}}{4\pi\epsilon_0}}\frac{q_{e}}{\hbar c},
\end{equation}
$V$ being the minimum speed and $m_{p}$ and $m_{e}$ are respectively the mass of the proton and electron. Such a minimum 
speed is $V=4.5876\times 10^{-14}$ m/s. We have found $\xi=1.5302\times 10^{-22}$\cite{11},
where Dirac's large number hypothesis (LNH) were taken into account in obtaining $\xi$, i.e.,
$F_e/F_g=q_e^2/4\pi\epsilon_0 G m_pm_e\sim 10^{40}$\cite{11}. 

It was shown\cite{11} that the minimum speed is connected to the cosmological constant in the following way:

\begin{equation}
V\approx\sqrt{\frac{e^{2}}{m_{p}}\Lambda^{\frac{1}{2}}}
\end{equation}.
\begin{figure}
\begin{center}
\includegraphics[scale=0.80]{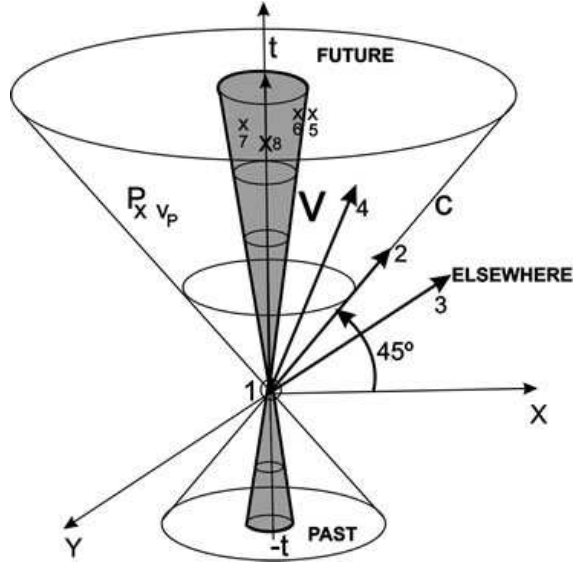}
\end{center}
\caption{The external and internal conical surfaces represent respectively the speed of light $c$ and the unattainable minimum
speed $V$, which is a definitely prohibited boundary for any particle. For a point $P$ in the world line of a particle, in the interior
of the two conical surfaces, we obtain a corresponding internal conical surface, such that we must have $V<v_p\leq c$. The $4$-interval
$S_4$ is a time-like interval. The $4$-interval $S_2$ is a light-like interval (surface of the light cone). 
The $4$-interval $S_3$ is a space-like interval (elsewhere). The novelty in spacetime of SSR are the $4$-intervals $S_5$ (surface of
the dark cone) representing an infinitly dilated time-like interval, including the $4$-intervals $S_6$, $S_7$ and $S_8$ inside the dark 
cone for representing a new space-like region (see ref.\cite{11}).}
\end{figure}

Therefore the light cone contains a new region of causality called {\it dark cone}\cite{11}, so that the speed of a particle
must belong to the following range: $V$(dark cone)$<v<c$ (light cone) (Fig.1). 

The breaking of Lorentz symmetry group destroys the properties of the transformations of Special Relativity (SR) and so generates
an intriguing kinematics and dynamics for speeds very close to the minimum speed $V$, i.e., for $v\rightarrow V$, we find new 
relativistic effects such as the contraction of the improper time and the dilation of space\cite{11}. In this new scenario,
the proper time also suffers relativistic effects such as its own dilation with respect to the improper one when 
$v\rightarrow V$, namely:

\begin{equation}
\Delta\tau\sqrt{1-\frac{V^{2}}{v^{2}}}=\Delta t\sqrt{1-\frac{v^{2}}{c^{2}}},
\end{equation}
which was shown in the reference\cite{11}, where it was also made experimental prospects for detecting such new relativistic 
effect close to the invariant minimum speed $V$, i.e., too close to the absolute zero temperature. 

Since the minimum speed $V$ is an invariant quantity as the speed of light $c$, $V$ does not alter the value of the speed $v$ of
any particle. Therefore we have called ultra-referential $S_{V}$\cite{10}\cite{11} as being the preferred (background) reference
frame in relation to which we have the speeds $v$ of any particle. In view of this, the reference frame transformations change 
substantially in the presence of $S_V$, as follows: 

a) The special case of $(1+1)D$ transformations in SSR\cite{7}\cite{8}\cite{9}\cite{10}\cite{11} with $\vec v=v_x=v$ (Fig.2) are 

\begin{equation}
x^{\prime}=\Psi(X-vt+Vt)=\theta\gamma(X-vt+Vt) 
\end{equation}

and 

\begin{equation}
t^{\prime}=\Psi\left(t-\frac{vX}{c^2}+\frac{VX}{c^2}\right)=\theta\gamma\left(t-\frac{vX}{c^2}+\frac{VX}{c^2}\right), 
\end{equation}
where $\theta=\sqrt{1-V^2/v^2}$ and $\Psi=\theta\gamma=\sqrt{1-V^2/v^2}/\sqrt{1-v^2/c^2}$.

\begin{figure}
\includegraphics[scale=0.80]{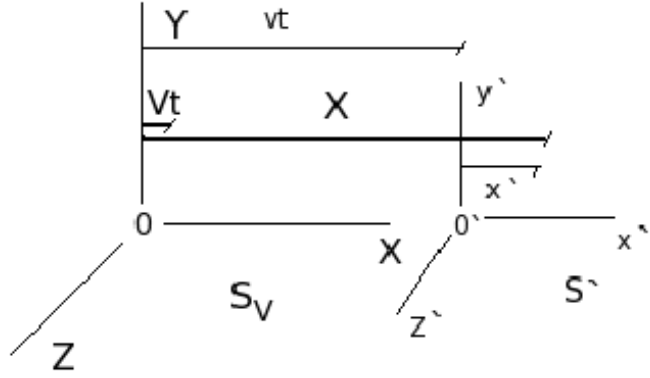}
\caption{In this special case of $(1+1)D$, the referential $S^{\prime}$ moves in $x$-direction with a speed $v(>V)$ with respect to the
 background field connected to the ultra-referential $S_V$. If $V\rightarrow 0$, $S_V$ is eliminated (empty space), and thus the galilean
 frame $S$ takes place, recovering Lorentz transformations.}
\end{figure}

b) The $(3+1)D$ transformations in SSR (Fig.3)\cite{11} are

\begin{equation}
\vec{r'}=\theta\left[\vec{r_{T}}+\gamma\left(\vec{r_{//}}-\vec{v}\left(1-\frac{V}{v}\right)t\right)\right]=
\theta\left[\vec{r_{T}}+\gamma\left(\vec{r_{//}}-\vec{v}t+\vec{V}t\right)\right]
\end{equation}

and

\begin{equation}
t'=\theta\gamma\left[t-\frac{\vec{r}\cdotp\vec{v}}{c^{2}}+\frac{\vec{r}\cdotp\vec{V}}{c^{2}}\right]. 
\end{equation}

Of course, if we make $V\rightarrow 0$, we recover the well-known Lorentz transformations.

\begin{figure}
\includegraphics[scale=0.80]{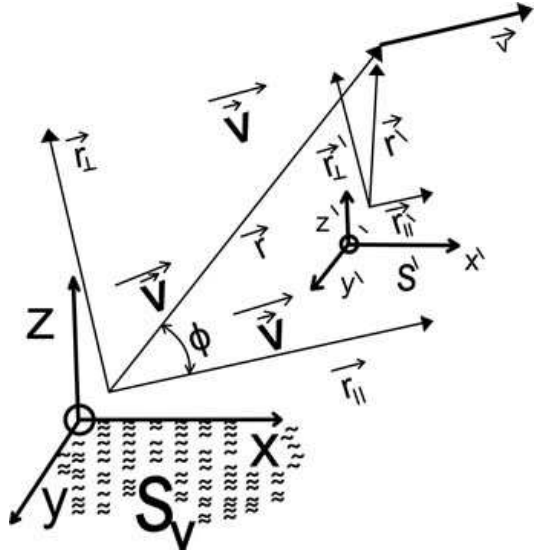}
\caption{$S^{\prime}$ moves with a $3D$-velocity $\vec v=(v_x,v_y,v_z)$ in relation to $S_V$. For the special case of $1D$-velocity
$\vec v=(v_x)$, we recover Fig.2; however, in this general case of $3D$-velocity $\vec v$, there must be a background vector $\vec V$
(minimum velocity) with the same direction of $\vec v$ as shown in this figure. Such a background vector $\vec V=(V/v)\vec v$ is 
related to the background reference frame (ultra-referential) $S_V$, thus leading to Lorentz violation. The modulus of $\vec V$ is invariant
at any direction.} 
\end{figure}

Although we associate the minimum speed $V$ with the ultra-referential $S_{V}$, this frame is inaccessible for any particle. Thus, the
effect of such new causal structure of spacetime generates an effect on mass-energy being symmetrical to what happens close to the speed 
of light $c$, i.e., it was shown that $E=m_0c^2\Psi(v)=m_0c^2\sqrt{1-V^2/v^2}/\sqrt{1-v^2/c^2}$, so that $E\rightarrow 0$ when
 $v\rightarrow V$\cite{10}\cite{11}. We notice that $E=E_0=m_0c^2$ for $v=v_0=\sqrt{cV}$\cite{11}. It was also shown that 
the minimum speed $V$ is associated with the cosmological constant, which is equivalent to a fluid (vacuum energy) with negative 
pressure\cite{10}\cite{11}.

The metric of such symmetrical spacetime of SSR is a deformed Minkowski metric with a global multiplicative function (a scale factor
with $v$-dependence) $\Theta(v)$ working like a conformal factor\cite{12}. Thus we write 

\begin{equation}
dS^{2}=\Theta\eta_{\mu\nu}dx^{\mu}dx^{\nu},
\end{equation}
where $\Theta=\Theta(v)=\theta^{-2}=1/(1-V^2/v^2)\equiv 1/(1-\Lambda r^2/6 c^2)^2$\cite{12},
working like a conformal factor and $\eta_{\mu\nu}$ is the Minkowski metric.

We can say that SSR geometrizes the quantum phenomena as investigated before (the origin of the Uncertainty Principle)\cite{9} in order to allow us to associate quantities belonging to the microscopic world with a new geometric structure that originates from Lorentz symmetry breaking. Such a geometry should be investigated in the future. 

\subsection{SSR-metric as a conformal metric in a DS-scenario and dS-metric} 

Let us consider a spherical universe with Hubble radius $r_u$ filled by a uniform vacuum energy density. On the surface of such a
sphere (frontier of the observable universe), the bodies (galaxies) experience an accelerated expansion (anti-gravity)
due to the whole ``dark mass (energy)" of vacuum inside the sphere. So we could think that each galaxy is a proof body interacting with
the big sphere of dark energy (dark universe) like in the simple case of two bodies interaction. However, we need to show that there is 
an anti-gravitational interaction between the 
ordinary proof mass $m_0$ and the big sphere with a ``dark mass" of vacuum ($M$). To do that, let us first start from the 
well-known simple model of a massive proof particle $m_0$ that escapes from a newtonian gravitational potential $\phi$
on the surface of a big sphere of matter, namely $E=m_0c^2(1-v^2/c^2)^{-1/2}\equiv m_0c^2(1+\phi/c^2)$, where $E$ is its relativistic
energy. Here the interval of escape velocity $0\leq v<c$ is associated with the interval of potential $0\leq\phi<\infty$, where we 
stipulate $\phi>0$ to be the attractive (classical) gravitational potential. 

Now we notice that Lorentz symmetry breaking due to the presence of the ultra-referential $S_V$ connected to the dark energy that 
fills the sphere has origin in a non-classical (non-local) aspect of gravity that leads to a repulsive gravitational potential
($\phi<0$). In order to see such an anti-gravity, let us consider the total energy of a proof particle on the surface of such a dark sphere according to SSR\cite{7}\cite{10}\cite{11}\cite{12}, namely:

\begin{equation}
E=m_0c^2\frac{\sqrt{1-\frac{V^2}{v^2}}}{\sqrt{1-\frac{v^2}{c^2}}}=
m_0c^2\left(1+\frac{\phi}{c^2}\right),
\end{equation}
from where we obtain
\begin{equation}
\phi=c^2\left[\frac{\sqrt{1-\frac{V^2}{v^2}}}{\sqrt{1-\frac{v^2}{c^2}}}-1\right], 
\end{equation}
where $m_0$ is the mass of the proof particle, $v$ being its input speed or also its escape velocity from the sphere. If the sphere is governed by
vacuum as occurs in the universe as a whole, then $v$ should be understood as an input speed in order to overcome anti-gravity, 
and thus the factor $\sqrt{1-V^2/v^2}$ prevails for determining the potential. However, for a sphere of matter, $v$ is the well-known 
escape velocity, so that the Lorentz factor takes place.  

From the above equation, we observe two regimes of gravitational potential, namely:

\begin{equation}
\phi= \left\{
\begin{array}{ll}
\phi_{Q}:&\mbox{$-c^2<\phi\leq 0$ for $V(=\xi c)< v\leq v_0$},\\\\
  \phi_{att}:&\mbox{$0\leq\phi<\infty$ for $v_0(=\sqrt{\xi}c=\sqrt{cV})\leq v< c$}, 
\end{array}
\right.
\end{equation}
where $v_0=\sqrt{cV}\sim 10^{-3}$m/s\cite{11}. 

$\phi_{att}$ and $\phi_{Q}$ are respectively the attractive (classical) and repulsive (non-classical or quantum)
potentials. We observe that the strongest repulsive potential is $\phi=-c^2$, which is associated with vacuum energy for the
ultra-referential $S_V$ (consider $v=V$ in Eq.(10)) (Fig.4).

\begin{figure}
\includegraphics[scale=0.8]{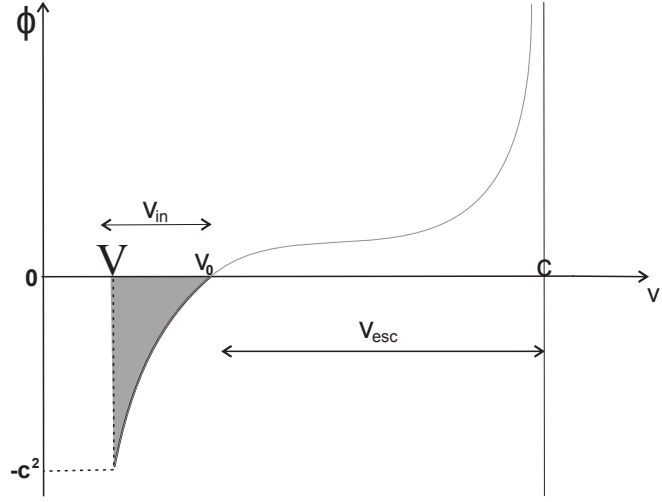}
\caption{This graph shows the potentials of SSR representing the function in Eq.(10) that presents two regimes, namely: a) The attractive (classical)
regime is well-known as Lorentz sector for describing gravity of a source of matter like a sphere of mass having a proof 
particle with mass $m_0$ on its surface. This particle escapes from this gravity with an escape velocity $v_0\leq v_{esc}<c$ according to
the attractive (positive) potential $0\leq\phi_{att}<\infty$. b) The repulsive (quantum) regime is the sector that provides the signature
of SSR for describing anti-gravity of a source of dark energy (vaccum energy) like an exotic sphere of dark mass having a proof 
particle of matter with mass $m_0$ on its surface. In this quantum sector, the escape velocity from anti-gravity should be understood as the 
input velocity $V<v_{in}\leq v_0$ according to the negative (quantum) potential $-c^2<\phi_{Q}\leq 0$, such that the proof particle with mass
$m_0$ is able to penetrate the dark sphere whose anti-gravity pushes it far away. Here we should observe that there is an  
intermediary velocity $v_0=\sqrt{cV}$, which corresponds to the point of phase transition between these two regimes in such a way 
that the general potential $\phi=0$. This means that $v_0$ can represent both escape and input velocities, which depends on the sector
we are considering. As we are just interested in the quantum sector (anti-gravity) of SSR, we have $v_{in}=v_0$ for $\phi=\phi_{Q}=0$ and
$v_{in}=V$ for $\phi=\phi_{Q}=-c^2$, since we just take into account the sector of negative potential for treating the extended
dS-relativity.}
\end{figure}

By considering the simple model of spherical universe with a radius $r_u$ and a uniform vacuum energy
density $\rho$, we find the total vacuum energy inside the sphere, i.e., $E_{\Lambda}=\rho V_u=-pV_u=Mc^2$, 
$V_u$ being its volume and $M$ the total dark mass associated with the dark energy.  Therefore, we are able to get a repulsive 
(negative) gravitational potential $\phi$ on the surface of such a sphere (universe) filled by dark ``mass'' (dark energy), namely: 

\begin{equation}
\phi=-\frac{GM}{r_u}=-\frac{G\rho V_u}{r_uc^2}=\frac{4\pi Gpr_u^2}{3c^2},
\end{equation}
where $p=-\rho$, $\rho$ being the vacuum energy density and $V_u=4\pi r_u^3/3$ (volume of the universe).

Knowing that $\rho=\Lambda c^2/8\pi G$, we write the repulsive potential as follows: 

\begin{equation}
\phi=\phi(\Lambda, r_u)=-\frac{\Lambda r_u^2}{6},
\end{equation}
from where, for any radius $r$ of the expanding universe, we generally write $\phi=-\Lambda r^2/6$\cite{11}. As this potential
represents the repulsive sector of gravity given in Eq.(10), we rewrite Eq.(10) by neglecting the Lorentz factor 
(sector of attractive gravity), and thus we obtain the approximation for the repulsive sector that includes all states of vacuum 
given by $v$ such that $V< v\leq v_0 (=\sqrt{\xi}c)$, namely:  

\begin{equation}
\frac{\phi}{c^2}=-\frac{\Lambda r^2}{6c^2}=\sqrt{1-\frac{V^2}{v^2}}-1, 
\end{equation}
such that, if $v=V$, we find $\phi(V)/c^2=-1$, so that $\phi(V)=-c^2$.  We have $-c^2<\phi<0$ (Fig.4). 

By manipulating Eq.(14), we can rewrite the scale factor $\Theta$ as follows:

\begin{equation}
\Theta=\frac{1}{\left(1-\frac{V^2}{v^2}\right)}=\frac{1}{\left(1+\frac{\phi}{c^2}\right)^2}
=\frac{1}{\left(1-\frac{\Lambda r^2}{6c^2}\right)^2}, 
\end{equation}
where we see that there are three equivalent representations for $\Theta$. 

Substituting Eq.(15) in Eq.(8), we write the spherical metric of SSR in the following way: 

\begin{equation}
d\mathcal S^{2}=-\frac{c^{2}dt^2}{\left(1-\frac{{\Lambda}r^2}{6c^2}\right)^2}+\frac{dr^{2}}{\left(1-\frac{{\Lambda}r^2}{6c^2}\right)^2}
+\frac{r^2(d\theta)^2+r^2\sin^2\theta(d\Phi)^2}{\left(1-\frac{{\Lambda}r^2}{6c^2}\right)^2}. 
\end{equation} 

We should note that $\phi/c^2=-\Lambda r^2/6c^2$, where we have $-c^2<\phi\leq 0$. So it is interesting to realize that, for the 
approximation $\phi>>-c^2$ or $\left|\phi_{Q}\right|<<c^2$, we are in the regime $v>>V$, which means a weakly repulsive regime that
corresponds to the present time of the universe whose temperature $T(\approx 2.73$K) connected to a certain velocity $v$ is still 
so far from $T=T_{min}(\sim 10^{-12}K)$\cite{13} connected to the minimum speed $V$. 

As Eq.(16) encompasses all types of vacuum, specially the ideal vacuum given for a too long time ($r\rightarrow r_{horizon}$) when the universe (vacuum energy density $\rho$) will be 
in a very strong repulsive regime with a very negative curvature 
$\mathcal R=-16\pi G\rho/c^2(1-V^2/v^2)\equiv -16\pi G\rho/c^2(1-\Lambda_{horizon} r^2/6c^2)^2\rightarrow -\infty$\cite{12}, now we can realize that only the approximation for a weakly repulsive regime is able to generate a special metric well-similar to DS-metric. So, in order to
see this special metric (like DS-metric) given only for weak anti-gravity, we just make the expansion of the denominator
of $\Theta$-factor in Eq.(16), so that we take into account only the first order term, since we are considering $\Lambda r^2/6<<c^2$
such that we have $(1-\Lambda r^2/6c^2)^2\approx (1-2\Lambda r^2/6c^2)=(1-\Lambda r^2/3c^2)$\cite{12}. Finally, in doing this approximation in SSR-metric [Eq.(16)], we find DS-metric, namely: 

\begin{equation}
d\mathcal S^{2}\approx dS^{2}_{DS}= -\frac{c^{2}dt^2}{\left(1-\frac{{\Lambda}r^2}{3c^2}\right)}+
\frac{dr^{2}}{\left(1-\frac{{\Lambda}r^2}{3c^2}\right)}+
\frac{r^2d\varOmega}{\left(1-\frac{{\Lambda}r^2}{3c^2}\right)}, 
\end{equation}
where $d\varOmega$ is   

\begin{equation}
 d\varOmega=(d\theta)^2+\sin^2\theta(d\Phi)^2
\end{equation}

We realize that the validity of DS-metric remains only in a weak anti-gravity regime as occurs in the case of the tiny positive cosmological constant given in the present time of the expanding universe. 

 \section{Deformed energy equation of a particle in Special Relativity due to the presence of a thermal background field}
 
According to the relativistic dynamics of Special Relativity (SR), the relativistic mass of a particle is $m=\gamma m_0$, where $\gamma=1/\sqrt{1-v^2/c^2}$ and
$m_0$ is its rest mass. On the other hand, according to Newton second law applied to its relativistic momentum, we find 
$F=dP/dt=d(\gamma m_0 v)/dt=(m_0\gamma^3)dv/dt=m_0(1-v^2/c^2)^{-3/2}dv/dt$, where $m_0\gamma^3$ represents an inertial mass 
($m_i$) that is larger than the relativistic mass $m(=\gamma m_0)$; i.e., we have $m_i>m$. 

The mysterious discrepancy between the relativistic mass $m$($m_r$) and the inertial mass $m_i$ from Newton second law is a controversial
issue\cite{14}\cite{15}\cite{16}\cite{17}\cite{18}\cite{19}\cite{20}. Actually the Newtonian notion about inertia as the resistance to 
acceleration ($m_i$) is not compatible with the relativistic dynamics ($m_r$) in the sense that we generally cannot consider 
$\vec F=m_r\vec a$. An interesting explanation for such a discrepancy is to take into consideration the influence of an isotropic background field that couples to the particle, by dressing its relativistic mass ($m_r$) in order to generate an effective
(dressed) mass $m^*(=m_{effective})$ working like the inertial mass $m_i(>m_r)$ in accordance with the Newtonian concept of inertia, where
we find $m^*=m_i=\gamma^2 m_r=\gamma^2 m$. In this sense, it is natural to conclude that $m^*$ has a nonlocal origin; i.e., it comes from
a kind of interaction with a background field connected to a universal frame\cite{11}, which is within the context of the ideas of  Sciama\cite{21}, Schr\"{o}dinger\cite{22} and Mach\cite{23}.

If we define the new factor $\gamma^2=\Gamma$, we write 

\begin{equation}
 m^*=\Gamma m,
\end{equation}
where $\Gamma$ provides a nonlocal dynamic effect due to the influence of a universal background field over the particle moving with
speed $v$ with respect to such a universal frame. According to this reasoning, the particle is not completely free, since its
relativistic energy is now modified by the presence of the whole universe, namely:  

\begin{equation}
E^*= m^*c^2=\Gamma m c^2
\end{equation}

As the modified energy $E^*$ can be thought as being the energy $E$ of the free particle plus an increment $\delta E$
of nonlocal origin, i.e., $E^*=\Gamma E=E+\delta E$, let us now consider that $\delta E$ comes from the thermal background field
of the whole expanding universe instead of simply a dynamic effect of a particle moving with speed $v$ in the background field,
in spite of the fact that there should be an equivalence between the dynamical and thermal approaches for obtaining the modified energy.
To show this, we make the following consideration inside the factor $\Gamma$, namely:

\begin{equation}
\Gamma(v)=\left(1-\frac{v^2}{c^2}\right)^{-1}\equiv\Gamma(T)=\left(1-\frac{\frac{m_Pv^2}{K_B}}{\frac{m_Pc^2}{K_B}}\right)^{-1}, 
\end{equation}
from where we find $\Gamma(T)=(1-T/T_P)^{-1}$, $T$ being the background temperature. $T_P(=m_Pc^2/K_B\sim 10^{32}$K) is the Planck
temperature in the early universe with Planck radius $R_P\sim 10^{-35}$m. $E_P(=m_Pc^2\sim 10^{19}$GeV) is
the Planck energy and $m_P(\sim 10^{-4}$g) is the Planck mass. From the thermal approach, if $T\rightarrow T_P$, $\Gamma(T)$ diverges. 

 Now we rewrite Eq.(20) as follows: 

   \begin{equation}
 E(T)=\Gamma(T)mc^2=\frac{\gamma m_0c^2}{1-\frac{T}{T_P}}. 
   \end{equation}

  As the factor $\Gamma(T)$ has a nonlocal origin and is related to the background temperature of the universe, let us admit that this 
factor acts globally on the speed of light $c$, while the well-known factor $\gamma$ acts locally on the relativistic mass of the
particle. In view of this, we should redefine Eq.(22) in the following way:

\begin{equation}
 E=[\gamma^{\prime} m_0][\Gamma(T)c^2]=\gamma^{\prime}m_0c^{\prime 2}=mc(T)^2=mc^{\prime 2},
\end{equation}
where now we have $m=\gamma^{\prime}m_0$, so that 

\begin{equation}
 \gamma^{\prime}=\frac{1}{\sqrt{1-\frac{v^2}{c^{\prime 2}}}}
\end{equation}

  And from Eq.(23) we extract

\begin{equation}
 c^{\prime}=c(T)=\frac{c}{\sqrt{1-\frac{T}{T_P}}}, 
\end{equation}
 where $c(T)=\sqrt{\Gamma(T)}c=\gamma_{T} c$, with $\gamma_{T}=1/\sqrt{1-T/T_P}$. 

From Eq.(25) we find that the speed of light was infinite in the initial universe when $T=T_P$. As the universe was expanding 
and getting colder, the speed of light had been decreased to achieve $c(T)\approx c$ for $T<<T_P$. Currently we have $c(T_0)=c$, with $T_0\approx 2.73$K.

\begin{figure}
\includegraphics[scale=0.40]{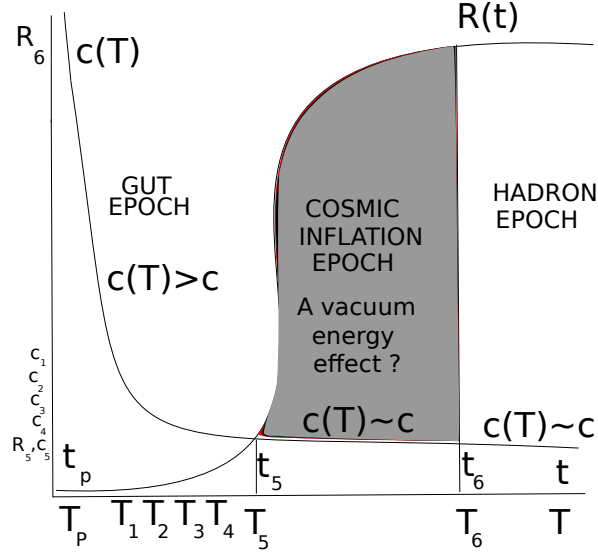}
\caption{This figure shows two graphics, namely $R(t)$, which is the size (radius) of the universe as a function of time, and $c(T)$,  
representing the speed of light with dependence on the temperature of the universe according to Eq.(25). At the beginning of
the universe when it was a singularity with a minimum radius of the order of the Planck radius, i.e., $R_P\sim 10^{-35}$m, 
having the Planck energy scale $E_p\sim 10^{19}$GeV which corresponds to the Planck temperature $T_P\sim 10^{32}$K and the Planck time
$t_P\sim 10^{-43}$s, the speed of light $c^{\prime}$ was infinite since there was no spacetime [see Eq.(25) for $c(T)$].
But immediately after, when $T_1=10^{31}$K, the speed of light had already assumed a value close to the current value as shown by
Eq.(25) for $c(T)$, and therefore a cone of light (a spacetime) had been formed; i.e., with $c=2.99792458\times 10^{8}$m/s for the
present time, then, according to the function $c(T)$, we find $c_1=c(T_1)=3.16008998\times 10^8$ m/s (see the figure). Subsequently,
for $T_2 =10^{30}$K, we find $c_2=c(T_2)=3.01302757\times 10^8$ m/s . For $T_3=10^{29}K\Rightarrow c_3=c(T_3)=2.99942467\times 10^8$m/s.
And for $T_4=10^{28}K\Rightarrow c_4= c(T_4)=2.99807447\times 10^8$m/s. Finally, for $T_5=10^{27}K\Rightarrow c_5=c(T_5)=
2.99793955\times 10^8$m/s. From this temperature $T_5=10^{27}$K, when $t=t_5\sim 10^{-35}$s, corresponding to the energy scale 
of the Grand Unified Theory (GUT) with $10^{14}$GeV, the universe inflated very quickly, starting with a radius $R_5\sim 10^{-25}$m and reaching $R_6\sim 10^{25}$m at the time $t_6\sim 10^{-32}$s; i.e., the size of the universe increased rapidly
50 orders of magnitude. Since the speed of light $c_5\approx c$, VSL should be put in doubt. Hence, perhaps the vacuum energy had played a fundamental role in that epoch. In the next
section, only when we will treat SSR with a thermal background field, this question about the vacuum energy in the inflationary period and also in the period of a far future time with 
accelerated expansion will be clarified.}
\end{figure}

The change in the speed of light is $\delta c=c^{\prime}-c$, namely:

\begin{equation}
 \delta c=(\gamma_{T}-1)c=\left(\frac{1}{\sqrt{1-\frac{T}{T_P}}}-1\right)c,
\end{equation}
where, for $T<<T_P$, we have $\delta c\approx 0$. 

 We should note that the variation of the speed of light with temperature does not invalidate the postulate of constancy of 
the speed of light in special relativity since $c^{\prime}$ for a given temperature remains invariant only with respect
to the motion of massive particles, but not with respect to the age and temperature of the universe.
In other words, we say that, although the speed of light has decreased rapidly during the initial expansion of 
the universe and thereafter with a smooth variation as shown in Fig.5, its value for a given temperature ($c(T)$) 
is still a maximum limit of speed that is invariant only with respect to the motion of all subluminal particles.

 The modified spatial momentum of a particle moving in the presence of a cosmic background field with temperature $T$ is the following: 

\begin{equation}
 P(T)=\gamma^{\prime}\gamma_{T} m_0v=\frac{\gamma^{\prime}m_0v}{\sqrt{1-\frac{T}{T_P}}}
\end{equation}

Before obtaining the modified energy-momentum relation, we first introduce the modified 4-velocity, namely: 

\begin{equation}
  U^{\prime\mu}=\left[\frac{\gamma^{\prime}c}{\sqrt{1-\frac{T}{T_P}}}~ , ~
\frac{\gamma^{\prime}v_{\alpha}}{\sqrt{1-\frac{T}{T_P}}}\right],
\end{equation}
where $\mu=0,1,2,3$ and $\alpha=1,2,3$. If $T\rightarrow T_P$, the 4-velocity diverges. 

The modified $4$-momentum is $P^{\prime\mu}=m_0U^{\prime\mu}$. So, from Eq.(28) we find

\begin{equation}
  P^{\prime\mu}=\left[\frac{\gamma^{\prime}m_0c}{\sqrt{1-\frac{T}{T_P}}}~ , ~
\frac{\gamma^{\prime}m_0v_{\alpha}}{\sqrt{1-\frac{T}{T_P}}}\right],
\end{equation}
where $E(T)=c^{\prime}P^{\prime 0}=m_0\gamma^{\prime}c^{\prime 2}=mc^2/(1-T/T_P)$. The 
spatial components ($\alpha=1,2,3$) of Eq.(29) represents the spatial momentum 
$P^{\prime}=P(T)=m_0\gamma^{\prime}v/\sqrt{1-T/T_P}$. 

From Eq.(29), by performing the quantity $P^{\prime\mu}P^{\prime}_{\mu}$, we obtain the modified energy-momentum relation as follows: 

\begin{equation}
 P^{\prime\mu}P^{\prime}_{\mu}=\frac{[E(T)]^2}{c^{\prime 2}}-[P(T)]^2 = m_0^2 c^{\prime 2},
\end{equation}
from which we get

\begin{equation}
 [E(T)]^2=\frac{m^2c^4}{\left(1-\frac{T}{T_P}\right)^2}=c^{\prime 2}[P(T)]^2 + m_0^2 c^{\prime 4},
\end{equation}
where $c^{\prime}=c/\sqrt{1-T/T_P}$ and $m=\gamma^{\prime}m_0$.

It is curious to notice that the Magueijo-Smolin doubly special relativity equation ($mc^2/1-E/E_P$)\cite{24} reproduces Eq.(22) when we just replace $E$ by $K_BT$ and $E_P$ by $K_BT_P$ in the denominator of their equation.  

\section{Energy equation of a particle in Symmetrical Special Relativity with the presence of a thermal background field}

Let us first consider a force applied to a particle, in the same direction of its motion. More general cases where the force is not 
necessarily parallel to velocity will be treated elsewhere. In our specific case ($\vec F||\vec v$), the relativistic power $P_{ow}(=vdp/dt)$ of SSR is given as follows:

\begin{equation}
P_{ow}=v\frac{d}{dt}\left[m_0v\left(1-\frac{V^2}{v^2}\right)^{\frac{1}{2}}\left(1-\frac{v^2}{c^2}\right)^{-\frac{1}{2}}\right],
\end{equation}
where we have used the momentum in SSR, i.e., $p=m_0v\Psi(v)$.

After performing the calculations in Eq.(32), we find

\begin{equation}
 P_{ow}=\left[\frac{\left(1-\frac{V^2}{v^2}\right)^{\frac{1}{2}}}{\left(1-\frac{v^2}{c^2}\right)^{\frac{3}{2}}}
+\frac{V^2}{v^2\left(1-\frac{v^2}{c^2}\right)^{\frac{1}{2}}\left(1-\frac{V^2}{v^2}\right)^{\frac{1}{2}}}\right]
\frac{dE_k}{dt},
\end{equation}
where $E_k=\frac{1}{2}m_0v^2$.

If we make $V\rightarrow 0$ and $c\rightarrow\infty$ in Eq.(33), we simply recover the power obtained in newtonian mechanics, namely
$P_{ow}=dE_k/dt$. Now, if we just consider $V\rightarrow 0$ in Eq.(33), we recover the well-known relativistic power of SR, namely
$P_{ow}=(1-v^2/c^2)^{-3/2}dE_k/dt$. We notice that such a relativistic power tends to infinite ($P_{ow}\rightarrow\infty$) in the
limit $v\rightarrow c$. We explain this result as an effect of the drastic increase of an effective inertial mass close to $c$, namely
$m_{eff}=m_0(1-v^2/c^2)^{k^{\prime\prime}}$, where $k^{\prime\prime}=-3/2$. We must stress that such an effective inertial mass is the
response to an applied force parallel to the motion according to Newton second law, and it increases faster than the relativistic 
mass $m=m_r=m_0(1-v^2/c^2)^{-1/2}$.

  The effective inertial mass $m_ {eff}$ that we have obtained is a longitudinal mass $m_L$, i.e., it is a response to the force applied in the
 direction of motion. In SR, for the case where the force is perpendicular to velocity, we can show that the transversal mass increases like the relativistic
 mass, i.e., $m=m_T=m_0(1-v^2/c^2)^{-1/2}$, which differs from the longitudinal mass $m_L=m_0(1-v^2/c^2)^{-3/2}$.  So, in this sense, there is 
anisotropy of the effective inertial mass to be also investigated in more details by SSR in a further work.

 In the previous section, it was already notice that the mysterious discrepancy between the relativistic mass $m$ ($m_r$) and the longitudinal inertial mass $m_L$ from Newton second law [Eq.(33)] is a controversial issue\cite{14}\cite{15}\cite{16}\cite{17}\cite{18}\cite{19}\cite{20}. Actually it is already known that the newtonian notion about inertia
 as the resistance to an acceleration
($m_L$) is not compatible with the relativistic dynamics ($m_r$) in the sense that we generally cannot consider $\vec F=m_{r}\vec a$. The 
 dynamics of SSR
 aims to give us a new interpretation for the inertia of the newtonian point of view in order to make it compatible with the relativistic
 mass. This compatibility is possible only due to the influence of the background field that couples to the particle and ``dresses" its relativistic mass in
 order to generate an
 effective (dressed) mass in accordance with the newtonian notion about inertia from Eq.(32) and Eq.(33). This issue will be clarified in this section.

 From Eq.(33), it is important to observe that, when we are closer to $V$, there emerges a completely new result (correction) for power, namely:

\begin{equation}
P_{ow}\approx\left(1-\frac{V^2}{v^2}\right)^{-\frac{1}{2}}\frac{d}{dt}\left(\frac{1}{2}m_0v^2\right),
\end{equation}
given in the approximation $v\approx V$. So, we notice that $P_{ow}\rightarrow\infty$ when $v\approx V$. We can also make the limit
$v\rightarrow V$ for the general case [Eq.(33)] and so we obtain an infinite power ($P_{ow}\rightarrow\infty$). Such a new relativistic effect deserves
the following very important comment:  Although we are in the limit of very low energies close to $V$, where the energy of the particle ($mc^2$) tends
to zero according to the approximation $E=mc^2\approx m_0c^2(1-V^2/v^2)^{k}$ with $k=1/2$, on the
other hand the power given in Eq.(34) shows us that there is an effective inertial mass that increases to infinite in the limit $v\rightarrow V$, that is to
say, from Eq.(34) we get the effective mass $m_{eff}\approx m_0(1-V^2/v^2)^{k^{\prime}}$, where $k^{\prime}=-1/2$. Therefore, from a dynamical point of view,
 the negative exponent $k^{\prime}$ ($=-1/2$) for the power at very low speeds [Eq.(34)] is responsible for the inferior barrier of the
 minimum speed $V$, as well as the exponent $k^{\prime\prime}=-3/2$ of the well-known relativistic power is responsible for the top barrier
 of the speed of light $c$ according to Newton second law. Actually, due to the drastic increase of $m_{eff}$ of a particle moving
 closer to $S_V$, leading to its strong coupling to the vacuum field in the background frame $S_V$, thus, in view of this, the dynamics 
 of SSR states that it is impossible to decelerate a subatomic particle until reaching the rest. This is the reason why there is a unattainable minimum speed $V$. 

 In order to see clearly both exponents $k^{\prime}=-1/2$ (inferior inertial barrier $V$) and $k^{\prime\prime}=-3/2$ (top inertial barrier
 $c$), let us write the general formula of power [Eq.(33)] in the following alternative way after some algebraic manipulations on it,
 namely:

\begin{equation}
P_{ow}=\left(1-\frac{V^2}{v^2}\right)^{k^{\prime}}\left(1-\frac{v^2}{c^2}\right)^{k^{\prime\prime}}\left(1-\frac{V^2}{c^2}\right)
\frac{dE_k}{dt},
\end{equation}
where $k^{\prime}=-1/2$ and $k^{\prime\prime}=-3/2$. Now it is easy to see that, if $v\approx V$ or even $v<<c$, Eq.(35) recovers the
approximation in Eq.(34). As $V<<c$, the ratio $V^2/c^2(<<1)$ in Eq.(35) is a very small dimensionless constant 
$\xi^2=V^2/c^2\sim 10^{-44}$\cite{11}. So $\xi^2$ can be neglected in Eq.(35).

So, from Eq.(35) we get the effective inertial mass $m^*=m_{eff}$ of SSR, namely:

\begin{equation}
m^*=m_{eff}=m_0\left(1-\frac{V^2}{v^2}\right)^{-\frac{1}{2}}\left(1-\frac{v^2}{c^2}\right)^{-\frac{3}{2}}. 
\end{equation}

 By taking into account the same reasoning as used before (Section 3) to interpret $m^*$
 within the context of a thermal background field, we also realize that the effective (inertial) mass $m^*$ has a nonlocal origin, which now presents a natural connection with SSR due to the existence of a preferred background frame related to the universal minimum speed\cite{11}. Thus SSR with a thermal background field will be able to predict the variation of the speed of light with temperature for a too far future cosmic time. 
 
 In order to obtain the variation of the speed of light for a very cold cosmological horizon close to a minimum temperature $T_{mim}(=m_PV^2/K_B\cong 3.28\times 10^{-12}K)$\cite{13}, 
 we first must get the factor $\Omega$ that transforms the inertial mass $m_0$ to the generalized inertial mass $m^*$ of SSR given in Eq.(36). So, let us now write the equivalent
 form of Eq.(36) in the following way: 
 
 \begin{equation}
   E^*=m^*c^2=\Omega m_0c^2=\left[\left(1-\frac{V^2}{v^2}\right)^{-\frac{1}{2}}\left(1-\frac{v^2}{c^2}\right)^{-\frac{3}{2}}\right]m_0c^2\equiv (\Gamma_{SSR}\Psi) m_0c^2,
  \end{equation}
where we already know that $\Psi m_0c^2=mc^2=E$. So we must stress that $E^*\neq E=mc^2$,
since $\Omega\neq\Psi$. As, in deformed SR, we have written $E^*=\Gamma mc^2=(\Gamma\gamma)m_0c^2$ with $\Gamma=\Gamma_{SR}=\gamma^2$, in an analogous way, from Eq.(37) we are able to obtain the factor $\Gamma_{SSR}$, so that we realize that 

\begin{equation}
 \Omega=\Gamma_{SSR}\Psi=\left(1-\frac{V^2}{v^2}\right)^{-\frac{1}{2}}\left(1-\frac{v^2}{c^2}\right)^{-\frac{3}{2}}.  
\end{equation}

As we have $\Psi=\sqrt{1-V^2/v^2}/\sqrt{1-v^2/c^2}$, from Eq.(38) we find 

\begin{equation}
 \Gamma_{SSR}=\Gamma_{SR}\left(1-\frac{V^2}{v^2}\right)^{-1}=\left(1-\frac{v^2}{c^2}\right)^{-1}\left(1-\frac{V^2}{v^2}\right)^{-1}. 
\end{equation}

In the same way that the deformation factor $\Gamma_{SR}$ is due to the presence of a thermal 
background field that increases significantly the energy $E$ of a particle only for higher temperatures close to Planck temperature ($T_P$) in the early universe, i.e., 
$E^*=\Gamma_{SR}E$ with $\Gamma_{SR}=\Gamma_{SR}(T)=(1-T/T_P)^{-1}$ and $E=mc^2$, the general deformation factor $\Gamma_{SSR}$ is also able to predict the
influence of a very cold thermal background field on the energy of a particle. 

According to a previous paper\cite{13}, we have demonstrated the existence of a universal 
minimum temperature $T_{min}(=m_PV^2/K_B\sim 10^{-12}K)$, which is related to a ultra-cold
cosmological horizon in a too far future time of a very old universe. We have shown 
the following thermal equivalence relation for lower energies which are associated to a ultra-cold background thermal bath, namely:

\begin{equation}
\left(1-\frac{V^2}{v^2}\right)^{-1}=\left[1-\frac{\left(\frac{m_PV^2}{K_B}\right)}
{\left(\frac{m_Pv^2}{K_B}\right)}\right]^{-1}\equiv\left(1-\frac{T_{min}}{T}\right)^{-1}. 
\end{equation}

So by substituting Eq.(21) and Eq.(40) in Eq.(39), we find the general deformation factor 
$\Gamma_{SSR}$ in its equivalent thermal form for representing the general effect of any thermal background field on the energy of a particle, as follows: 

\begin{equation}
 \Gamma_{SSR}(T)=\left(1-\frac{T_{min}}{T}\right)^{-1}\left(1-\frac{T}{T_P}\right)^{-1}.
\end{equation}

By substituting Eq.(41) in Eq.(37), we finally obtain the deformed energy $E^*$ of a
particle in SSR due to the temperature of the universal background field, namely: 

\begin{equation}
 E(T)=\Gamma_{SSR}(T)mc^2=\frac{\Psi m_0c^2}{\left(1-\frac{T_{min}}{T}\right)\left(1-\frac{T}{T_P}\right)}. 
\end{equation}

If we make $V\rightarrow 0$ in Eq.(42), this implies $T_{min}\rightarrow 0$, and thus Eq.(22)
is recovered as a special case just for higher background temperatures of an early 
universe governed by the cosmic inflation. 

Before continuing to investigate the important implications of Eq.(42), it is interesting to note that SSR without temperature ($E=\Psi m_0c^2$) remains valid for a long period of cosmic time when $T_{min}<<T<<T_P$.  

In a similar reasoning that was used to interpret the factor $\Gamma_{SR}(T)$ in Eq.(22) as 
being of non-local origin by acting globally on the speed of light $c$, then the general factor $\Gamma_{SSR}(T)$ in Eq.(42) should be also interpreted as being of non-local origin by acting on the speed of light $c$, since its non-local aspect is due to the background temperature of the whole universe. Thus let us admit that this thermal factor acts globally on the speed of light $c$, while the kinematics factor $\Psi$ acts locally on the relativistic mass of the particle. In view of this, we should redefine Eq.(42) as follows: 

 \begin{equation}
 E(T)=[\Psi^{\prime} m_0][\Gamma_{SSR}(T)c^2]=\Psi^{\prime}m_0c^{\prime 2}=mc(T)^2=
 mc^{\prime 2},
\end{equation}

where we have $m=\Psi^{\prime}m_0$, so that we write 

\begin{equation}
 \Psi^{\prime}=\frac{\sqrt{1-\frac{V^{\prime 2}}{v^2}}}{\sqrt{1-\frac{v^2}{c^{\prime 2}}}},  
\end{equation}
where $c^{\prime}=c(T)$ will be obtained soon and $V^{\prime}=V(T)$ will be obtained in the next section. 

\begin{figure}
\includegraphics[scale=0.80]{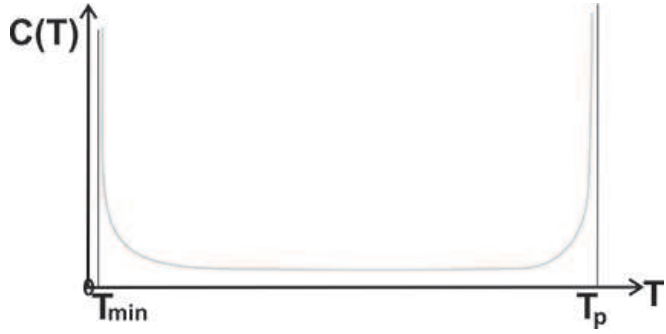}
\caption{This graph for representing Eq.(45) shows that the speed of light diverges for both
limits of temperature, namely Planck temperature $T_P(\sim 10^{32}K)$ for the scale of 
Planck $L_P(\sim 10^{-35}m)$ in the early (too hot) universe, and a minimum temperature $T_{min}(\sim 10^{-12}K)$ in a ultra-cold universe connected to a horizon radius $r_h>>r_u(\sim 10^{26}m)$.} 
\end{figure}

By inserting Eq.(41) into Eq.(43), we finally extract $c(T)$, namely: 

 \begin{equation}
c^{\prime}=c(T)=\frac{c}{\sqrt{1-\frac{T_{min}}{T}}\sqrt{1-\frac{T}{T_P}}}.
 \end{equation}
 
 In Eq.(45), if we make $T_{min}\rightarrow 0$, we recover Eq.(25) that represents the 
 particular case of $c(T)$ in the deformed SR with the presence of a background thermal field
 as it was well investigated in the previous section where we have shown that the drastic decreasing of the speed of light for $T<T_P$ is not able to explain the horizon problem
 (background isotropy) in the hot universe (Fig.5). This result calls into question the VSL theories that counteract the inflationary model that aims to explain the background isotropy (horizon problem), although the inflationary model does not still provide a clear explanation for the origin of the cosmic inflation field usually so-called {\it ``inflaton''}. 
 
 It is very curious to notice that Eq.(45) provides a similarity between the too hot
 universe ($T\approx T_P$) close to the Planck scale $L_P$ and the ultra-cold universe 
 ($T\approx T_{min}$) close to the horizon radius ($r_h=\sqrt{6}c/\sqrt{\Lambda_h}$\cite{12}) in the sense that the speed of light diverges for both limits. This leads us to think that there emerges another inflation very close to the ultra-cold horizon ($r_h$), i.e., there emerges a very rapid stretching of a ultra-cold space whose temperature is too close to $T_{mim}$, which is responsible for the drastic increasing of the speed of light being dragged by very cold inflation itself. This novelty is provided only by SSR with the presence of a thermal background field. In this sense, the theory shows that both inflationary periods, i.e., the initial inflation and the final rapid accelerated expansion,
 have the same origin related to a too hot and cold vacuum respectively. 
 
  The current vacuum energy is related to the well-known cosmological constant $\Lambda_0\sim 10^{-35}s^{-2}$, but in a very far future time, the theory predicts that the temperature will decrease until it will approach $T_{mim}$ when the cosmological constant will also decrease until approaching to a horizon cosmological constant $\Lambda_h$ for $r\rightarrow r_h$. So a new inflation will begin due to the appearance of an infinitely negative curvature, i.e., there will emerge a Big Rip of the spacetime tissue for a 
  ultra-cold vacuum, thus leading to a very rapid increasing of the speed of light that will
  instantly illuminate the whole exploding universe. To show all these effects in such a limit of ultra-cold universe, we should realize that there is an equivalence between Eq.(15) and Eq.(40), so that we find 
  
  \begin{equation}
 c^{\prime}=\frac{c}{\sqrt{1-\frac{T_{min}}{T}}}\equiv\frac{c}{\left(1-\frac{\Lambda_h r^2}{6c^2}\right)},    
  \end{equation}
where we can see that, for $T\rightarrow T_{min}$ or for $r\rightarrow r_h$, the speed of
light diverges so rapidly that a ultra-cold inflation of the whole universe begins. At a 
first sight, such an inflation seems to lead to the so-called {\it Big Rip} of space-time tissue, however, as the minimum temperature $T_{min}$ and the horizon radius $r_h$ are both unattainable, there could be strong fluctuations of vacuum during the rapid expansion so that
the temperature could drastically fluctuates in many small parts of space, thus leading to an enourmous number of very hot inflationary bubbles that would emerge of such parts and thus
many expanding ``baby universes'' working like bubbles would be created from the final inflation and so on. Then, in a distant past, probably our universe energed from drastic fluctuations of a small part in the order of Planck scale in the scenario of a previous expanding universe (``mother universe''). This a reasonable conjecture according to the present theory, and it is in a certain accordance with some other theories about ``mother universes'' and ``baby universes''. However, according to such conjectures, some puzzles arise, namely: Was there an uncreated primordial universe from which all the mother and baby universes have been arisen? And if such a primordial ``universe'' is the first cause of all others, then how did this first creation take place? These intriguing questions are still on hold. In the next section, we intend to go deeper into to these questions with the present theory. 

  According to a previous paper\cite{12}, we have obtained the curvature $\mathcal R$ of an
extended DS-space governed by all kind of vacuum (cosmological constants), where the fundamental vacuum is given by the minimum speed related to a horizon cosmological constant.
So, now taking into account Eq.(40) and Eq.(15), we can write $\mathcal R$\cite{12} in the 
following way:

\begin{equation}
\mathcal R =-\frac{16\pi G\rho}{c^2\left(1-\frac{T_{min}}{T}\right)}\equiv -\frac{16\pi G\rho}{c^2\left(1-\frac{\Lambda_h r^2}{6c^2}\right)^2}. 
\end{equation}

 If $T\rightarrow T_{min}$ or $r\rightarrow r_h$, this implies that the scalar curvature
 of the universe governed only by ultra-cold vacuum becomes infinitely negative,i.e., 
 $\mathcal R\rightarrow -\infty$. Such an infinitely negative curvature is responsible 
 by an infinite anti-gravity that stretches drastically the space by dragging the light
 so that its speed increases to the infinite. In this sense, we realize that there is 
 a direct connection between the scalar curvature and the variation of the speed of light 
 with temperature of the expanding universe close to the cosmological horizon. To show this,
 we just compare Eq.(46) with Eq.(47), and so we find
 
 \begin{equation}
  \mathcal R (T)=-\frac{16\pi G\rho}{c^4}[c(T)]^2,
 \end{equation}
 where $c(T)=c\sqrt{1-T_{min}/T}$. 
 
 According to Eq.(45), the change in the speed of light is $\delta c=c^{\prime}-c$, namely:

\begin{equation}
 \delta c=(\Gamma_{SSR}-1)c=\left(\frac{1}{\sqrt{1-\frac{T_{min}}{T}}
 \sqrt{1-\frac{T}{T_P}}}-1\right)c,
\end{equation}
where, for $T_{min}<<T << T_P$, we have $\delta c\approx 0$.

In order to obtain the modified energy-momentum relation in SSR with a thermal background
field, let us introduce the modified $4$-velocity, namely, 

\begin{equation}
\mathcal U^{\prime\mu}=\left[\frac{\Psi^{\prime}c}{\sqrt{1-\frac{T_{min}}{T}}
\sqrt{1-\frac{T}{T_P}}}~ , ~
\frac{\Psi^{\prime}v_{\alpha}}{\sqrt{1-\frac{T_{min}}{T}}\sqrt{1-\frac{T}{T_P}}}\right],
\end{equation}
where $\mu=0,1,2,3$ and $\alpha=1,2,3$. If $T\rightarrow T_P$ or even $T\rightarrow T_{min}$, 
the $4$-velocity diverges. 

The modified $4$-momentum is $\mathcal P^{\prime\mu}=m_0\mathcal U^{\prime\mu}$. So, 
from Eq.(50) we obtain 

\begin{equation}
  \mathcal P^{\prime\mu}=\left[\frac{\Psi^{\prime}m_0c}{\sqrt{1-\frac{T_{min}}{T}}\sqrt{1-\frac{T}{T_P}}}~ , ~\frac{\Psi^{\prime}m_0v_{\alpha}}{\sqrt{1-\frac{T_{min}}{T}}\sqrt{1-\frac{T}{T_P}}}\right],
\end{equation}
where $E(T)=c^{\prime}\mathcal P^{\prime 0}=m_0\Psi^{\prime}c^{\prime 2}=
mc^2/(1-T_{min}/T)(1-T/T_P)$.

From Eq.(51),by performing the quantity $\mathcal P^{\prime\mu}\mathcal P^{\prime}_{\mu}$, we find the following modified energy-momentum relation as follows: 

\begin{equation}
 \mathcal P^{\prime\mu}\mathcal P^{\prime}_{\mu}=\frac{[E(T)]^2}{c^{\prime 2}}-[P(T)]^2=
 m_0^2 c^{\prime 2}\left(1-\frac{V^{\prime 2}}{v^2}\right),
\end{equation}
from which we get

\begin{equation}
 [E(T)]^2=\frac{m^2c^4}{\left(1-\frac{T_{min}}{T}\right)^2\left(1-\frac{T}{T_P}\right)^2}=c^{\prime 2}[P(T)]^2 + m_0^2 c^{\prime 4}\left(1-\frac{V^{\prime 2}}{v^2}\right),
\end{equation}
where we have the spatial momentum of deformed SSR, namely $P(T)=\Psi^{\prime}m_0 v/\sqrt{(1-T_{min}/T)}\sqrt{(1-T/T_P)}$ and $c^{\prime}=c/\sqrt{1-T_{min}/T}\sqrt{1-T/T_P}$ [Eq.(45)].  

 Eq.(53) represents the dispersion relation of deformed SSR with the presence of a thermal 
 background field in the cosmological scenario of the expanding universe, where both  inflationary primordial and final epochs are taken into account.

 \section{Variation of the universal minimum speed in both scenarios of an inflationary universe and final accelerated expansion} 
 
 \subsection{The concept of reciprocal velocity in SSR: the uncertainty on position}

As already discussed in a previous paper\cite{9}, SSR generates a kinematics of non-locality as also proposed in the emergent gravity theories\cite{25}. 

In order to see more clearly the aspect of non-locality of SSR due to the stretching of space when $v\approx V$, we should take into account the idea of the so-called reciprocal velocity $v_{rec}$, which has been already well explored in a previous paper\cite{9}. Thus, here we will just reintroduce such idea in a more summarized way. To do that, let us use Eq.(3) and first multiply it by $c$ at both sides, and after by taking the squared result in order to obtain   

\begin{equation}
 \left(c^2-\frac{c^2V^2}{v^2}\right)(\Delta\tau)^2=(c^2-v^2)(\Delta t)^2, 
\end{equation}
where the right side of Eq.(54) related to the improper time $\Delta t$ provides the velocity $v$ of particle ($\Delta t\rightarrow\infty$ for $v\rightarrow c$) and, on the other hand, the left side gives us the new information that shows that the proper time in SSR is not an invariant quantity as is in SR, so that the proper time goes to infinite ($\Delta\tau\rightarrow\infty$) when $v\rightarrow V$, which leads to a too large stretching of the proper space interval $c\Delta\tau$ in this limit of much lower speed, giving us the impression that the particle is well delocalized due its ``high internal speed" that is so-called reciprocal velocity that appears at the left side of the equation as being $v_{rec}=(cV)/v=v_0^2/v$. So we have $(c^2-v_{rec}^2)(\Delta\tau)^2=c^2-v^2)(\Delta t)^2$. Now we can perceive that the reciprocal velocity $v_{rec}$ represents a kind of inverse of $v$ such that, when $v\rightarrow V$, we get $v_{rec}\rightarrow c$, i.e., the ``internal motion'' is close to $c$, thus leading to the new effect of {\it proper time dilation} associated to a delocalization that was shown as being an uncertainty on position\cite{9} within
the scenario of spacetime in SSR. In this scenario, it was shown that the decreasing of momentum close to zero ($v\approx V$) leads to a delocalization of the particle, which is justified by the increasing of $v_{rec}\rightarrow c$ and the dilation of the proper 
time $\Delta\tau\rightarrow\infty$. As the uncertainty on position is $\Delta x=v_{rec}\Delta\tau=(v_0^2/v)\Delta\tau$\cite{9}, for $v\rightarrow V$, we find $\Delta x=c\Delta\tau\rightarrow\infty$. And, on the other hand, the large increasing of momentum for $v\rightarrow c$ leads to the well-known dilation of the improper time $\Delta t$ (right side of Eq.(54)), so that we find $\Delta\tau<<\Delta t$ and the minimum reciprocal velocity $v_{rec}\rightarrow v_0^2/c=V$, which provides a small uncertainty on position, since 
$\Delta x=V\Delta\tau$.

Therefore Eq.(54) or Eq.(3) can be rewritten in the following way: 

\begin{equation}
\Delta\tau\sqrt{1-\frac{v^2_{rec}}{c^{2}}}=\Delta t\sqrt{1-\frac{v^{2}}{c^{2}}},
\end{equation}
where we find $v^2_{rec}/c^2=V^2/v^2$. We have $V<v<c$ and $V<v_{rec}<c$, where $V$ is the reciprocal of $c$ and vice-versa. 

\begin{figure}
\includegraphics[scale=0.60]{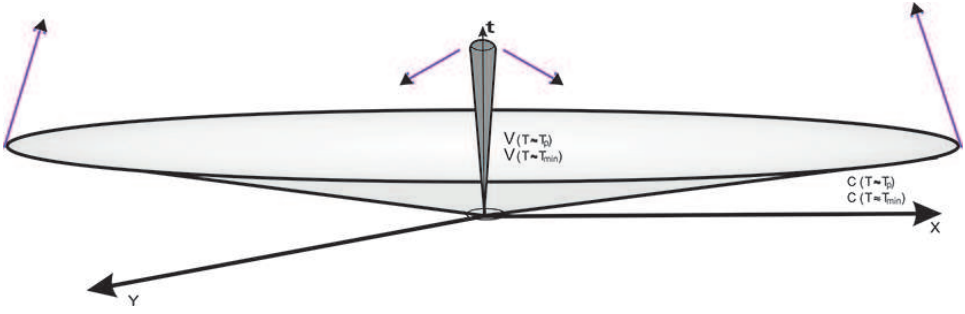}
\caption{The figure shows that the dark cone and light cone are opposite aspects of a same
transcent state working like a Newtonian space where $c^{\prime}\rightarrow\infty$ and 
$V^{\prime}\rightarrow 0 $ for both limits of temperatures, namely $T_P$ and $T_{min}$.} 
\end{figure}

As the minimum speed is the reciprocal of $c$, i.e., we have $V=v_0^2/c$, then we simply obtain the $V(T)$, namely:

\begin{equation}
 V(T)=\frac{v_0^2}{c(T)}, 
\end{equation}
where we must call attention to the fact that $v_0=\sqrt{c(T)V(T)}=\sqrt{cV}$ is a universal fixed point that does not have dependence on temperature because it represents the unique 
point where occurs the phase transition between gravity (ADS-space with positive curvature
for $v>v_0$ or $\Phi>0$ with $\Lambda<0$ ) and anti-gravity (DS-space with negative curvature for $v<v_0$ or $\Phi<0$ with $\Lambda>0$), according to Eq.10 given in Fig.4. Thus $v_0$ is the perfect Newtonian regime where the curvature is exactly null, i.e., it is a perfectly flat space where temperature does not make sense, however such a point do not have stability in any spacetime, since any spacetime is necessarily the result of the existence of two barriers given by a dark cone for a certain minimum limit $V^{\prime}=V(T)$ and a light cone for certain maximum limit $c^{\prime}=c(T)$ (see Fig.7). Otherwise, it would not be a physical space. In this sense based on SSR, we conclude that a Newtonian space would merely be a non-physical idealization of a perfectly inertial ``space" (see letter {\it a} in Fig 8 representing such flat space). 

  By introducing Eq.(45) for $c(T)$ into Eq.(56), we obtain its reciprocal speed $V(T)$ 
  (Fig.7), namely: 
  
\begin{equation}
V(T)=V\sqrt{1-\frac{T_{min}}{T}}\sqrt{1-\frac{T}{T_P}},
\end{equation}
where it is easy to verify that $v_0$ is in fact a fixed (invariant) Newtonian point, since
we get $c(T)V(T)=cV=v_0^2$ when multiplying Eq.(45) by Eq.(57). 

Both Eq.(45) and Eq.(57) show respectively that $c(T_P)$ and $c(T_{min})$ diverge, while 
$V(T_P)$ and $V(T_{min})$ vanish (no dark cone). Of course, the absence of the dark cone 
would be the result of the absence of a light cone as the speed of light becomes infinite, so that there would be no light to cast darkness, and so this dialectical (dual) idea of
thesis X anti-thesis based on a dynamical symmetry would be overcame by an absolute permanent 
state for representing the non-physical (Newtonian) flat space without temperature or even
at a zero absolute temperature ($T=0K$).

Figure 7 shows clearly an abysmal gap from a non-physical state (a flat space or a space without fluctuations: see letter {\it a} of Fig.8) to a dynamical spacetime where the emergence of temperature (vibration) leads to the finite speeds of light and the non-zero minimum speeds until reaching their well-known current values that generate our expanding spacetime.

To summarize, SSR theory makes us rethink deeply that such an ethernal dialectical process of creation, expansion, and destruction of universes is sustained by an even more fundamental permanent (Newtonian) state from which there was an abysmal leap that provided a perturbation in the flat space (the letters {\it b} and {\it c} in Fig.8) and thus a bubble on the Newtonian absolute empty space has emerged for representing our universe (letter {\it d} in Fig.8), although a multiverse may also arise mysteriously from this non-physical Newtonian state, which seems to be a First Cause since there are no fluctuations (letter {\it a} in Fig.8). The misterious passage between such a Newtonian absolute state (uncreated primordial ``universe'') and the spacetime seems to bring back a transcendent aspect that surpasses the dialectical materialism of the modern cosmology of the cyclic universes, where such an ideal state of Newtonian (flat) space associated to a fixed point $v=v_0=\sqrt{c(T)V(T)}=\sqrt{cV}$ is still completely neglected by the idea of dialectical materialism.

\begin{figure}
\includegraphics[scale=0.80]{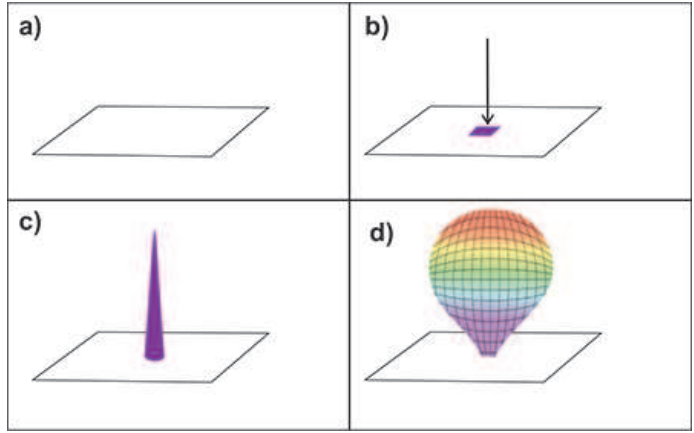}
\caption{A Newtonian or flat space works like an uncreated primordial universe. For an unknown reason, this ``serene lake" (null curvature) is in the eminence of being disturbed. An infinite negative curvature arises generating a vacuum with a very strong anti-gravity which creates a high peak at the Planck scale. The temperature that increases drastically leads to the emergence of an inflationary bubble that will generate our universe.} 
\end{figure}

\section{Conclusions and prospects}

We already know that the superfluid generated by SSR is associated with the cosmological constant\cite{11}. In the future, such a relationship may allow us to address problems associated with gravitational collapse. The study of the symmetries of SSR must also 
be done soon with the calculation of the symmetries and their association with a new kind of electromagnetism when we are in the limit $v\rightarrow V$, which could explain the problem of high magnetic fields in magnetars\cite{26}, super-fluids in the interior of gravastars and other kinds of black hole mimickers so-called quantum black holes.


\begin{thebibliography}{99}

\bibitem{1} A. Albrecht and J. Magueijo, Phys. Rev. D{\bf 59}, 043516 (1999).
\bibitem{2} J. Magueijo, Phys. Rev. D{\bf 62}, 103521 (2000). 
\bibitem{3} J. Magueijo, Phys. Rev. D{\bf 63}, 043502 (2001).
\bibitem{4} J. Magueijo, Rept. Prog. Phys. {\bf 66}, 2025 (2003). 
\bibitem{5} J. Magueijo , {\it Faster Than the Speed of Light-The Story of a Scientific
Speculation}, (Perseus Books,Boston,MA, 2003), ISBN 0-7382-0525-7. 
\bibitem{6} J. Casado, arXiv:astro-ph/0310178.
\bibitem{7} C. Nassif, Pramana Journal of Physics, Vol.71, 1, p.1-13 (2008).
\bibitem{8} C. Nassif, {\it Deformed Special Relativity with an energy barrier of a minimum speed}, International Journal of Modern Physics D Vol.19, No.5, p.539 (2010).
\bibitem{9} C. Nassif, {\it Doubly Special Relativity with a minimum speed and the Uncertainty Principle}, International Journal of Modern Physics D, Vol.21, N.2, p.1-20 (2012). 
\bibitem{10} C. Nassif: {\it An explanation for the tiny value of the cosmological constant and the low vacuum energy density}, General Relativity and Gravitation Vol.47, 9, p.1-34 (2015).
\bibitem{11} C. Nassif: {\it ``On the electrodynamics of moving particlesin a quasi flat spacetime with Lorentz violation and its cosmological implications''}, International Journal of
Modern Physics D Vol.25, 10 (2016) 1650096 (67 pages) \\
see in OPEN ACCESS: http://www.worldscientific.com/worldscinet/ijmpd?journalTabs=read
\bibitem{12} C. Nassif, R. F. dos Santos and A. C. Amaro de Faria Jr.:{\it ``Lorentz violation
with a universal minimum speed as foundation of de Sitter relativity''}, International Journal
of Modern Physics D Vol. 27 (2018) 1850011 (22 pages)\\
see in OPEN ACCESS: http://www.worldscientific.com/worldscinet/ijmpd?journalTabs=read
\bibitem{13} C. Nassif, A. C. Amaro de Faria Jr. and R. F. dos Santos:{\it ``Testing Lorentz symmetry violation with an invariant minimum speed''}, Modern Physics Letters A Vol. 33, No. 23 (2018) 1850148 (15 pages)\\
https://www.worldscientific.com/doi/10.1142/S0217732318501481 
\bibitem{14} C. J. Adler, Am. J. Phys.{\bf 55}, 739 (1987). 
\bibitem{15} R. P. Feynman , R. B. Leighton and M. Sands, {\it The Feynman Lectures on Physics} (Addison-Wesley,Reading, MA, 1963), Vol.1.
\bibitem{16} V. L. Okun , Physics Today {\bf 42} No.6, 31 (1989).
\bibitem{17} T. R. Sandin, Am. J. Phys. {\bf 59}, 1032 (1991). 
\bibitem{18} W. Rindler, {\it Introduction to Special Relativity} (Clarendon Press, Oxford 1982),pp. 79-80. 
\bibitem{19} W. Rindler, {\it Essential Relativity} (Springer, New York, 1977), 2nd ed. 
\bibitem{20} E. F. Taylor and J. A. Wheeler, {\it Spacetime Physics} (W.H.Freeman, New York 1992), 2nd ed.,pp. 246-251.
\bibitem{21} D. W. Sciama, Mon. Not. R. Astron. Soc.{\bf 113}, 34 (1953). 
\bibitem{22} E. Schr\"{o}dinger, Ann. Phys. (Berlin) {\bf 382}, 325 (1925). 
\bibitem{23} E. Mach,{\it The Science of Mechanics} (Open Court, La Salle, 1960). 
\bibitem{24} J. Magueijo and L. Smolin, Phys. Rev. Lett. 88, 190403 (2002).
\bibitem{25} D. Marolf, in:{\it Emergent Gravity requires (kinematics) non-locality},
10.1103/Phys.Rev.Lett.114.031104.
\bibitem{26} J. Greiner {\it et al}, Nature 523, 189-192 (2015).
\end{thebibliography}
\end{document}